# Quantum billiards with correlated electrons confined in triangular transition metal dichalcogenide monolayer nanostructures created by laser quench.

Jan Ravnik[1,2], Yevhenii Vaskivskyi[1], Jaka Vodeb[1,3], Polona Aupič[1], Igor Vaskivskyi[1], Denis Golež[4,5], Yaroslav Gerasimenko[6], Viktor Kabanov[1], Dragan Mihailovic[1,6]

[1]Dept. of Complex Matter, Jožef Stefan Institute, Jamova 39, 1000 Ljubljana, Slovenia.
[2]Laboratory for Micro and Nanotechnology, Paul Scherrer Institut, Forschungsstrasse 111, 5232 Villigen PSI, Switzerland
[3]Dept. of Physics, Faculty of Mathematics and Physics, University of Ljubljana, Jadranska 19, 1000 Ljubljana, Slovenia
[4]Dept. of Theoretical Physics, Jožef Stefan Institute, Jamova 39, 1000 Ljubljana, Slovenia.
[5]Flatiron Institute, New York, New York, USA
[6]CENN Nanocenter, Jamova 39, 1000 Ljubljana, Slovenia

**Forcing systems though fast non-equilibrium phase transitions offers the opportunity to study new states of quantum matter that self-assemble in their wake. Here we study the quantum interference effects of correlated electrons confined in monolayer quantum nanostructures, created by femtosecond laser-induced quench through a first-order polytype structural transition in a layered transition-metal dichalcogenide material. Scanning tunnelling microscopy of the electrons confined within equilateral triangles, whose dimensions are a few crystal unit cells on the side, reveals that the trajectories are strongly modified from free-electron states both by electronic correlations and confinement. Comparison of experiments with theoretical predictions of strongly correlated electron behaviour reveals that the confining geometry destabilizes the Wigner/Mott crystal ground state, resulting in mixed itinerant and correlation-localized states intertwined on a length scale of 1 nm. Occasionally, itinerant-electron states appear to follow quantum interferences which are suggestive of classical trajectories (quantum scars). The work opens the path toward understanding the quantum transport of electrons confined in atomic-scale monolayer structures based on correlated-electron-materials.**

A laser-induced quench through a first order structural transition can create small domain structures with atomically precise shapes that are beyond the reach of current nanofabrication technologies. Viewed from a quantum physics perspective, such structures represent a fruitful playground for investigating the quantum behaviour of particles in geometrically confined systems. Regular shapes, and equilateral triangles (ETs) in particular, are of special interest because they allow the study of the crossover from periodic limit cycles to chaotic trajectories[1–3]. Quantum scars – quantum interference (QI) patterns that follow traces of the paths of classical particles[4–7] - were investigated until now in fabricated mesoscopic semiconductor heterostructures and graphene by scanning tunnelling microscopy[8,9]. However, with strongly interacting electrons such patterns are not expected due to their tendency for localization. Recently, oscillating patterns were observed in interacting Rydberg atom arrays [exp] and theoretically motivated by the existence of many-body scars - manifold of low entangled excited states[10,11]. Correlation effects may be expected to give rise to perturbed trajectories in which the entanglement dynamics shows a dependence on the nature of the perturbation. The observation of QI in confined correlated electron systems would be of both fundamental interest, and also of practical importance for designing coherent electron devices with correlated materials.

Here we use scanning tunnelling microscopy to investigate QI in ET-shaped monolayer nanostructures of $TaS_2$ as small as ~2.6 nm to ~12.5 nm wide (8 ~ 38 unit cells) (Fig. 1). $TaS_2$ is a prototypical electronically correlated quasi-2D material, which is prone to carrier localization and the formation of different charge orders at different temperatures that become commensurate at 'magic' filling fractions[12]. The orthorhombic 1T polytype of 1T-$TaS_2$ below ~180 K, has a commensurate charge-density-wave (CCDW) with a large modulation amplitude of ~ 1 electron per 13 Ta sites, localized in a $\sqrt{13} \times \sqrt{13}$ superlattice structure, which can exhibit either left (L) or right (R) chirality with respect to the crystal lattice[13] (Fig. 1a). The lattice surrounding each 13$^{th}$ Ta atom is distorted by the extra charge [14] resulting in the formation of a polaron. Due to Coulomb interactions between such polarons, the system is correlated and thought to be susceptible to the formation of a Mott state[14,15], whence it is often discussed in terms of a polaronic Wigner crystal[12,16,17]. The 2H (trigonal) polytype is metallic above 75 K, but forms a commensurate CCDW below this temperature, which – in contrast to the 1T polytype – is metallic down to 1 K, below which it becomes superconducting[18].

The ET nanostructures are created by a laser pulse-induced quench through an inversion-symmetry-breaking polytype transformation of the surface atomic monolayer of a 1T polytype $TaS_2$ single crystal. The resulting ET domains are embedded laterally by a 1H-$TaS_2$ crystalline layer (the 1H signifies it is a monolayer). The entire structure is epitaxial on a 1T-$TaS_2$ single crystal (Fig. 1 a,b)).

Understanding the behaviour of correlated electrons confined in such small ETs theoretically represents a substantial challenge. Considering CDWs as standing wave interferences formed by counter-propagating Fermi electrons[19–21] suggests an investigation using the QB approach with Fermi electrons. On the other hand, the 2D polaronic Wigner crystal picture, where the polarons are subject to Mott localization[14] suggests a correlated electron picture. Here we compare conventional QB calculations, a (classical) charged lattice gas strongly correlated electron model, and a fully quantum many-body correlated electron model using exact diagonalization methods with the aim to understand the rich variety of QI textures in both classical and quantum regimes observed by STM.

## Experimental measurements and analysis

The ET structures (Fig. 1f) are created by a controlled exposure of a freshly exfoliated 1T-TaS$_2$ single crystal to laser pulses in ultrahigh vacuum at 80 K, where the majority of the top surface is transformed to the 1H polytype, but ET structures of 1T polytype remain structurally unchanged.[22].. The domains have atomically defined sides parallel to the crystal axes of the 1T layer, matching the lattice structure of the surrounding 1H layer, forming a perfect ET shape with edges at 60° to each other (Fig. 1 a, b). The work functions for the 1T-TaS$_2$ and 2H-TaS$_2$ polytypes are $\phi = 5.6$ and $5.2\ eV$ respectively, so the H phase acts as a barrier for electrons in the 1T ET nanostructure[23]. At 80K, where the measurements are preformed, the 1H polytype is metallic, while the 1T polytype is nominally in the insulating CCDW phase[14]. The H-polytype thus defines a confining potential barrier $\phi_B$ to the electrons inside the ET (Fig. 1e). As a result of self-organization of charges at the $1T - 1H$ interface, an edge state is formed which is clearly visible in ETs of all sizes (Fig. 1f). An interfacial band diagram based on a conventional metal-semiconductor junction is shown in Fig. 1e (see SI for details). The width $w$ of the edge state (ES) is approximately equal to the screening length in 1T-TaS$_2$, $\zeta \simeq 2\ nm$ [24,25].

The presented STM images in Fig. 1f and Fig. 2 are measurements of the local density of states (LDOS), which may be considered either in the local state approximation as $\rho_{local}(E, \boldsymbol{r})$ at the tip position $\boldsymbol{r}$, for states without long-range translation invariance, $\rho_{local}(E, \boldsymbol{r}) \propto \Sigma_{i=1,N} |\psi_i(E_i, \boldsymbol{r})|^2 \delta(E - E_i)$; or in the quasiparticle approximation $\rho_{QP}(E, \boldsymbol{r}) \propto \Sigma_k |\psi_k(r)|^2 \delta(E - \varepsilon(\boldsymbol{k}))$, where $\varepsilon(\boldsymbol{k})$ is the energy of all the electrons with different wavevector $\boldsymbol{k}$ that interfere locally at position $\boldsymbol{r}$. The latter case is relevant when we discuss the electrons in the unconfined 1H layer. It applies also to the interfering itinerant ET-confined electron eigenstates with different $\boldsymbol{k}_i$ but the same energy $\boldsymbol{E}(\boldsymbol{k})$. In either case, the resulting LDOS patterns inside the ET can be observed by STM topography at constant current. The QI is superimposed on the fine sub-nm structure of the atomic orbitals (see, for example, the high-resolution image in the insert to Fig. 1 f). This detailed orbital structure of surface atoms is related to structural effects and is not of present concern, so we shall focus primarily on the mesoscopic QI patterns.

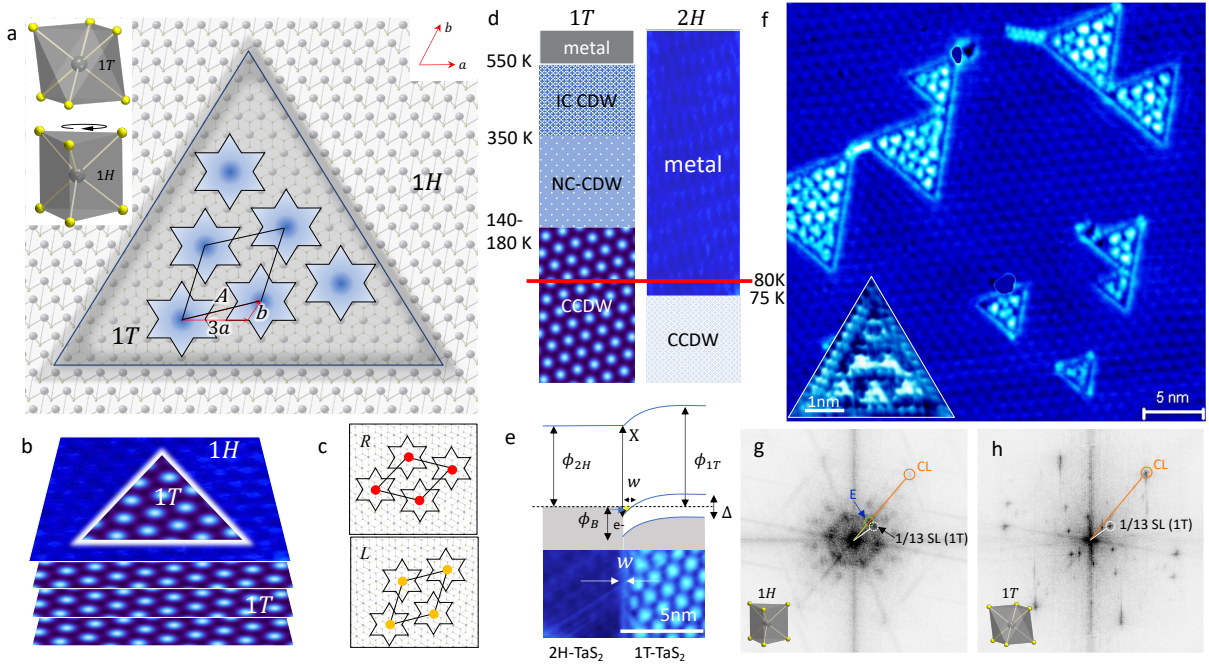

Figure 1. Monolayer 1T-TaS$_2$ structure in the shape of an ET bounded by a 1H-TaS$_2$ monolayer. a) A schematic picture of the left-handed packing of polarons in a C superlattice within an ET of outer dimensions 16a (inner dimensions l=11a), where a=0.33 Å is the lattice constant. The CCDW superlattice vector ***A*** is shown in terms of the lattice vectors **a** and **b**. (Fig. 1a). The unit cells of the 1T and 1H crystal structures are shown in the insert. b) The schematic structure of ETs on top of a 1T-TaS$_2$ substrate. c) Ideal polaron packing in L and R-handed CCDW superlattices. d) A phase diagram showing the charge density wave ordering transitions of the 1T- and 2H-polytypes as a function of temperature. The measurement temperature and phase transition temperatures are indicated. e) A band diagram of the 1T-1H boundary. An edge state of width w forms within the 1T phase as a result of band bending (see SI). f) An STM image of different sized a 1T-TaS$_2$ *ETs* embedded laterally within a 1H-TaS$_2$ monolayer on the surface of a 1T-TaS$_2$ single crystal. Note the ubiquitous presence of the edge state. The insert (bottom-left) shows a high-resolution image of a small ET with 6 polarons. g,h) Fourier transforms of the 1H and 1T regions respectively. The 1/13 CCDW SL and CL peaks are indicated. A peak attributed to the polarons ordered parallel to the edge of the ETs is also indicated (E).

Large-area Fourier transforms of the inside and outside of the ETs, shown in Figs. 1g and 1h respectively, reveal both crystal lattice (CL) peaks and CCDW superlattice (SL) peaks. The areas outside the ETs give sharp FFT peaks corresponding to the 1H CL and weak SL peaks from the 1T-TaS$_2$ CCDW layer underneath (Fig. 1h). Inside the ETs, we see an additional intensity that is almost uniformly distributed between the CCDW FFT peaks. This is attributed to QI features, and the periodic ordering along the inside edges of the ETs (labelled E).

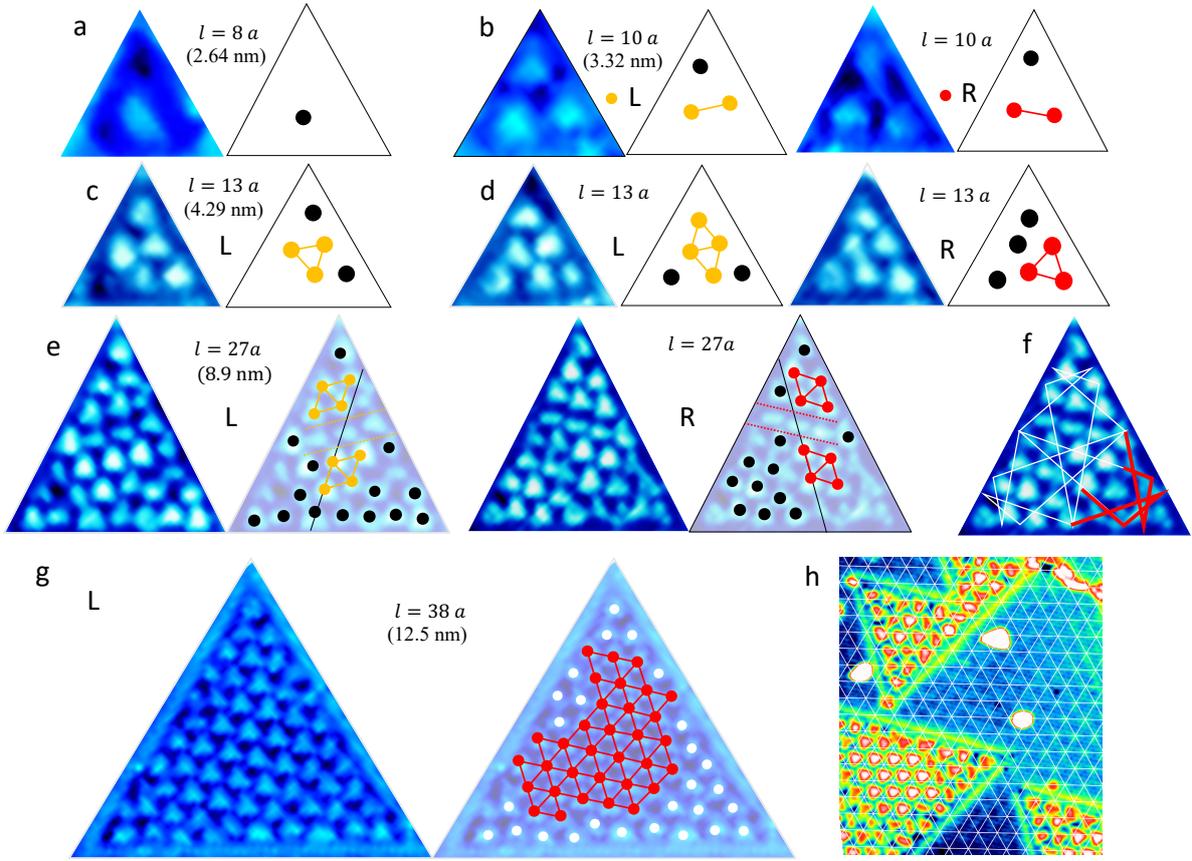

Figure 2 STM images of different sized ETs with l = 8...38a at 80K with a bias voltage of 0.8V. Each STM image is accompanied by a schematic figure indicating the appearance of L or R chirality of the C order (yellow and red respectively). Multiple localized QI features do not conform to the C order (black dots). a) A single dot is observed for l=8a. b) For l=10a, L and R chiralities of C order appear. c) l=13a with 5 polarons, d) l=13a with 6 polarons of L and R chiralities. e) l=27a with L and R chiralities created nearby with the same laser pulse exposure. Similar domain wall patterns (dashed lines) are observed in both cases, as schematically shown. f) l=27a R, but showing a superimposed classical periodic orbit with integers m=1, n=4 defining the 'shooting' vector **V** of motion of the classical particle, $V = ma + nb$ on the parent lattice (see ref. [26] for details). The red line emphasizes a reginal similarity of the classical trajectory with the QI pattern. g) An ET with l=38a showing a single C domain at the center, surrounded by QPI at the edges. h) the registry of the CCDW orders in the ETs and the 1T-TaS₂ layer below. The grid pattern is centered on the CCDW in the 1T layer below the 1H layer. All ETs show a phase shift of the polaron order with respect to the substrate, and each other.

In Fig. 2 we show a representative set of STM images of the inside of ETs, with inner ET dimensions ranging from $l = 8a$ (2.64 nm) to $38a$ (12.35 nm), where $a$ is the CL constant of 1T-TaS$_2$. Universally, we observe that the LDOS texture tries to adapt to the ET boundaries *and* form a commensurate order at the center. The CCDW superlattice structure is significantly distorted, however, particularly in small ETs. For the smallest ETs such as the one with $l = 8\ a$ (Fig. 2a), a single, deformed dot is visible at the center of the ET. Already with $l = 10\ a$ (Fig. 2b), the pattern with 3 maxima adjusts to the preferred CCDW order of either L or R chirality. In Fig. 2 c, d, we clearly see very different QI textures in ETs with the same $l$, which is likely caused by small geometrical imperfections and/or initial conditions that results in different electron trajectories within ETs of equal size. In the same vein, two ETs with $l =$ $27a$ (Fig. 2f) created simultaneously nearby to each other show quite complicated but closely matching, mirror images of opposite chirality[27]. A remarkable feature of these ETs, particularly well visible in the R structure in Fig. 2e is the non-trivial QI pattern in the corner. This does not fit the CCDW order and is strongly suggestive of a quantum interference that follows a predicted trajectory of classical particles within an ET[26] (shown by red line in Fig. 2f., and chosen to fit to the QI pattern). As the size of the ETs further increases, the polaron pattern approaches the CCDW superstructure. Thus, for $l = 38a$ the CCDW fills almost the entire ET, with QI distortions visible only at the edges.

Remarkably, the CCDW charge order within the ETs is *not* in register with respect to the CCDW order of the layer below. (The latter is visible through the surrounding 1H monolayer, and is emphasized by the mesh in Fig.2h.) Moreover, neighbouring ETs also have different register (Fig. 2 f). This implies that the effect of the underlying CCDW and lattice potential is not sufficiently strong to force the CCDW order in the top layer. The QI patterns appear to be determined by the ET boundary conditions, not by the inter-layer coupling as has been suggested in bulk crystals[28].

Tunnelling spectra

The tunnelling spectra inside, outside and across the edge of an ET at 80K are presented as the normalized differential conductance (NDC), $(dI/dV)/(I/V)$ in Fig. 3b. The recorded positions are color coded. For comparison we also show a 1T-TaS$_2$ CCDW bulk crystal spectrum at 4K showing the characteristic upper and lower Hubbard bands (UHB and LHB) at $+0.15$ and $-0.22$ V, and CCDW-derived bands at $+ 0.28$ and $-0.42$ V[29].

Overall, the states within $\pm 0.32$ $eV$ of the Fermi level correspond to Ta-5$d$ zone-folded sub-bands of the 1T-TaS$_2$ CCDW phase[30], and pristine Ta-5d bands of the 1H monolayer which give rise to the observed NDC. A most striking feature is the spatial homogeneity of the NDC curves *outside* the ET and the edge state. This is contrasted by the huge spatial spectral variations inside the ET occurring on a scale of 1 nm. Outside the ETs, beside the broad bands, a small, asymmetric gap-like feature is visible around $\pm 20$ mV with peaks at $\pm 50$ mV. This feature is similar to the one reported in single layer 1H-TaSe$_2$ on epitaxial bi-layer graphene substrates at 5 K [31] and monolayer 2H-TaS$_2$ on graphene [18], where it is was attributed to the CDW gap of the 1H monolayer. Considering that the data were recorded at 80 K, this would suggest that in the monolayer, the CDW gap is already present above the $T_c = 75$ K of the bulk material. However, no modulation is visible in the FFTs corresponding to the $3 \times 3$ period of the 1H-TaS$_2$ CCDW (Fig. 1g), implying that there is no long-range CDW order in the 1H layer. In some positions we also observe signatures of the characteristic peak at +0.25 V that corresponds to the UHB of bulk 1T-TaS$_2$, which are attributed to the CCDW in the layer below.

The ES also shows remarkable homogeneity along both sides of the boundary (green and pink dots respectively). The NDCs on the outside edge (the bright border, green) are very similar to the 1H monolayer (yellow). However, the broad peak at -0.21V (yellow) splits into two, at -0.15V and -0.25V (green). Inside the ET, the ES peak at -0.25 V shifts further to -0.28 V (pink), but no other significant differences are observed.

Inside the ET, the NDC curves vary significantly. A number of sharp peaks are observed that have no parallel in bulk 1T or 2H-TaS$_2$. However, large spatial variations of LDOS are indeed expected for a confined system with no long-range order. For example, the observation of nodal domains in space is one of the expected features for a QB[32] that modify the CCDW pattern. The sharp features on the energy scale $\pm 0.1\ V$ are thus attributed to eigenstates of the confined nanostructure.

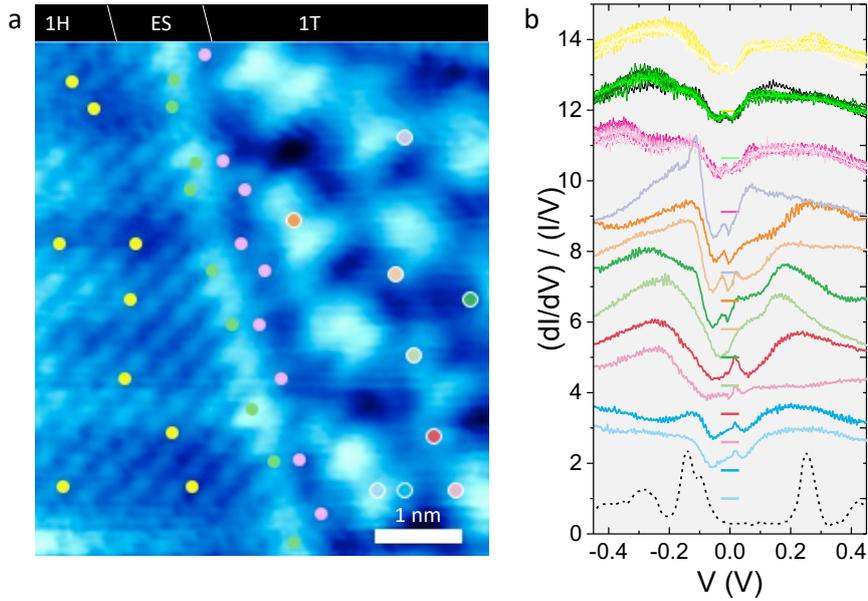

Figure 3. STS curves of the 1H layer outside, the 1T layer inside, and the ES of an ET. a) An STM image with indicated positions where STS curves are recorded, using color coded dots. b) The STS curves (in corresponding color) for the 1H layer, the ES <u>inside</u>, and <u>outside</u> the ET boundary (indicated by a narrow line), and the 1T phase inside an ET nanocrystal. The zero level is indicated in each case by a dashed line in the same color code. The STS of the CCDW state of a 1T-TaS$_2$ monocrystal is shown for comparison (dotted).

## Discussion

Within a non-interacting picture, electrons confined within ETs are expected to exhibit canonical QB behaviour. On the other hand, the presence of strong correlations localizes electrons in a commensurate structure[12], filling all available space. The observed LDOS patterns imply that the electron trajectories within the ETs are strongly modified by correlations, forcing the electrons along paths that make the probability density commensurate with the CCDW.

To investigate this behavior in more detail theoretically, we first compare the expected LDOS patterns within ETs using a traditional QB model, introducing an effective potential to signify the presence of a periodic lattice distortion accompanying the CCDW in the layer below, that may force electronic order in the top 1T layer. Then, introducing correlations in the classical limit, we calculate the expected localization patterns using a sparsely filled charged lattice gas model[12]. Finally, in an attempt to describe the itinerant electrons subject to correlations, we calculate electron density patterns that show spatial localization textures using exact diagonalization methods.

**The quantum billiard with inter-layer interactions.** For an ideal ET with multiple non-interacting electrons, the levels are filled subject to the Pauli principle. The calculated LDOS, $\rho_{local}(E, \mathbf{r})$ based on solutions of the Schrödinger equation are compared with the STM images for different ETs in Fig. 4, column A). The LDOS is given by $\Sigma_{N=n..m}|\psi_N|^2$, where the integers $n$ and $m$ indicate the range of eigenstates for the summation. By choosing appropriate $n$ and $m$ (by inspection), the predicted LDOS patterns show the correct number of maxima within the ET, but the pattern is always symmetric with respect to the ET shape, and parallel with the edges, which the experimental patterns are not. While qualitatively describing the LDOS patterns, the QB approach fails to describe the unusual features of the experimentally observed LDOS (column D of Fig. 4.)

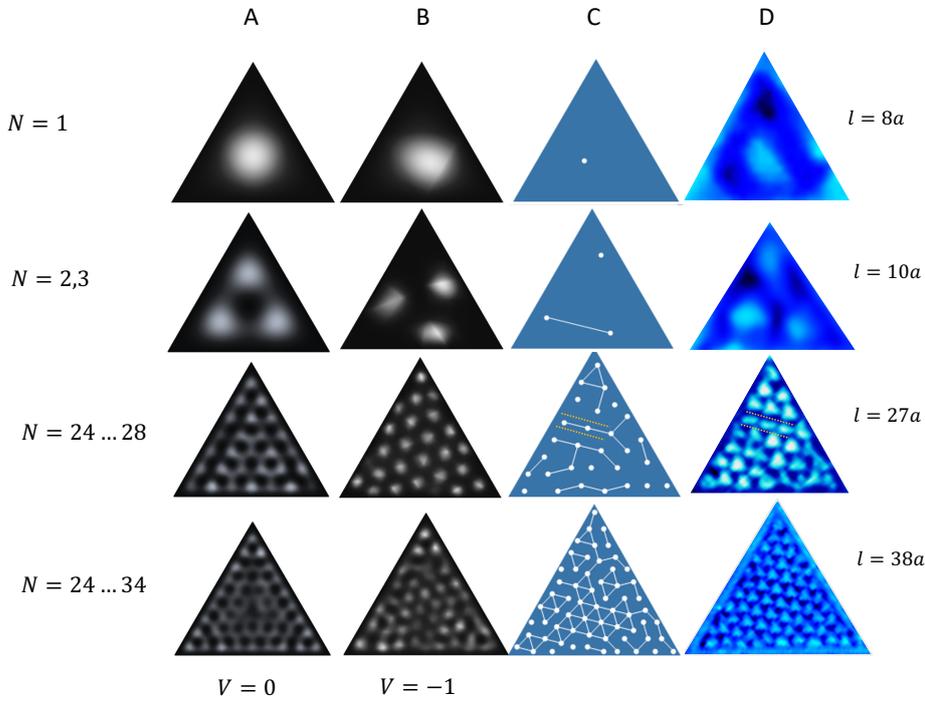

Figure 4 A comparison of QB, QB+V and CLG model predictions for $\rho_{local}(E, \mathbf{r})$ with experiment for ETs of different sizes with different sides $l$. Column A represents the QB solution (V=0) for different summed states (N), corresponding to $V_B \simeq 0.8V$. Column B shows the QB+V model eigenstates with $= -1$, and column C is the CLG MC calculation result at 1/13 filling. Column D shows experimental images for $l = 8\,a$, $10\,a$, $27\,a$ and $38\,a$. Note the domain wall predicted for $l = 27a$ (col. C), and corresponding STM image (col. D) shown by dashed lines.

The energy scale of the states is given by the confinement energy $E \sim \frac{N^2 \hbar^2 \pi^2}{2m^* l^2}$. Assuming $m^* \sim m_e$, for the smallest ET with side $l = 2.15$ nm ($8a$), the first four levels $N = 1 \ldots 4$ have energies ranging from $E \approx 0.08 - 1.26 eV$, while in the largest ET shown here, with $l = 12.8\,nm$ ($38a$), for $N = 1 \ldots 21$, $E \approx 0.0022 - 0.97 eV$. A more realistic effective mass $\sim 3\,m_e$ would compress the energy scale, increasing the number of experimentally observed levels in the STS range $\pm 0.5\ V$. The temperature broadening at 80 K is ~7 meV, so the levels are likely too closely spaced to be experimentally resolved. Nevertheless, the energy scale of the observed spectral features (Fig. 3b) is in qualitative agreement with the experiment.

Next, we introduce a periodic potential $V(x, y)$ in the QB calculation, $-\frac{\hbar^2}{2m}[\Delta_{x,y} + V(x,y)]\psi(x,y) = E\psi(x,y)$, to account for the periodic lattice distortion created by interaction of the top layer with the CCDW in the layer below. The results are shown in Column B of Fig. 4. Now the 2D pattern corresponding to the CCDW is tilted at 13° with respect to the ET edges, and the crystal axes. As expected, for a significantly large $V(r)$, the calculated LDOS pattern follows the CCDW ordering at the center of the ETs quite well. The periodic potential also introduces a gap in the LDOS which bears resemblance to the observed STS gap (Fig. 3). However, the calculation fails to reproduce the complex features and malleable nature of the LDOS patterns at the edges. It also completely fails to reproduce the observed domain walls. (The predicted $\rho_{local}(E, r)$ patterns and LDOS spectra for the QB, and QB+V models are described in the SI).

**Correlated electrons confined within an ET.** A correlated electron model is needed to account for the malleable features in the electronic order within ETs. Within a charge-lattice gas (CLG) Monte-Carlo (M-C) calculation, classical point charges, subject to screened Coulomb repulsion can move via thermal hopping on an atomic lattice shaped in the form of an ET (Fig 4c). A crucial parameter in the presented modelling is the filling, defined as the number of electrons divided by the number of lattice sites $f$. At various magic fillings, such as $f = \frac{1}{13}$ for the case of bulk 1T-TaS$_2$, the model predicts an electronic superlattice which is perfectly commensurate with the underlying atomic lattice[12]. Confining the system size to a small triangle inevitably introduces edge effects and distortions into the configurational ordering of electrons. This is simply due to the fact that a $\frac{1}{13}$ electronic lattice does not match the edges of the triangle as shown in Figure 4c, requiring the particles to accommodate. (For more details concerning the model refer to the SI.) Here we outline the similarities between experimental observations and the simulated electronic configurations (Fig. 4, column C): (1) The model predicts the spontaneous formation of two chiralities of the 1/13 electronic lattice at angles $\pm 13.9°$ with respect to the ET edges, which are both experimentally observed. In contrast to previous calculations where the external potential fixed the angle of the QI pattern, here it emerges as a nontrivial consequence of many-body correlations. (2) Smaller triangles induce stronger edge effects into the electron configuration, to the point of entirely breaking up the expected 1/13 electronic lattice. Electrons at the triangle's edge align with the edge and since the edges are close one to another, there is no room for a proper 1/13 lattice to emerge. The resulting configuration is still largely influenced by strong correlations, as electrons on average are $\sqrt{13}$ atomic lattice spacings apart. This is shown by the connecting lines in Figure 4 (column C). (3) Edge effects in larger triangles are diminished towards the centre, as shown for $l = 38a$ in Figure 4 (column D), and the 1/13 Wigner crystal lattice emerges. However, near the edges the electrons still align with the edge of the ET.

In the simple classical calculation above, all charges are equivalent and should ideally have the same LDOS and thus appear identical under a tunnelling microscope. However, the experimental STM images show various irregular, elongated and triangular shapes within the triangles that are sometimes aligned in rows (e.g. Fig. 4) which such modelling cannot describe.

For an understanding of the departures from the simple CLG model, we need to consider the itinerant correlated electrons which follow quantum billiard trajectories inside the triangle. To account for the interplay of the strong electronic repulsions and their itinerant nature we perform a quantum many-body calculation on small ET using the exact diagonalization of the spin-less fermions with the long-range interaction:

$$H = -t \sum_{\langle i,j \rangle} c_i^\dagger c_j + \sum_{i \neq j} V(|i-j|)[n_i - \overline{n}][n_j - \overline{n}],$$

where $c_i$ ($n_i$) is the annihilation (density) operator at site $i$, $t$ is the hopping parameter, V(|i−j|) is the Yukawa interaction. To ensure charge neutrality, we have subtracted the uniform background charge density n (see Methods for details). While in the classical limit the relevant parameter is filling $N$, the quantum extension introduces another dimensionless parameter $t/V$ that governs the transition between the localized ($t/V = 0$) and the delocalized regime ($t/V \gg 1$). In agreement with the classical simulation, there exist special fillings where the electrons form commensurate fillings of the ET. Due to the size restrictions in the many-body simulation, we consider an ET with the lattice size $L = 8$ and the number of electrons $N = 10$ corresponding to the special filling f=1/3, see Fig. 5 a). In this commensurate situation, the state is extremely robust against the delocalization which can be observed by comparing the density distribution in the quantum $t/V = 0$ and the classical case $t/V = 0.01$, see Fig. 5 a). The analogous comparison of the density distribution for the incommensurate filling exhibits a strong redistribution of charges forming nontrivial QPI and their delocalization tendency is driven by lowering the kinetic energy via the closed loops QPI, see Figs. 5 c) and d).

As a more quantitative measure of the electron delocalization with respect to the classical limit we introduce the parameter $D = \sqrt{\Sigma_i [\langle n_i \rangle - \langle n_i \rangle^{t=0}]^2}$. A comparison for different fillings shows that incommensurate states ($N \neq 10$) are orders of magnitude more susceptible to delocalization than commensurate ones (Fig. 5 e). The geometrical constrains of ET can therefore dramatically shift the delocalization transitions and this resolves the apparent contradiction how we can observe delocalized QPI patterns in small structures while the corresponding bulk situation would be localized.

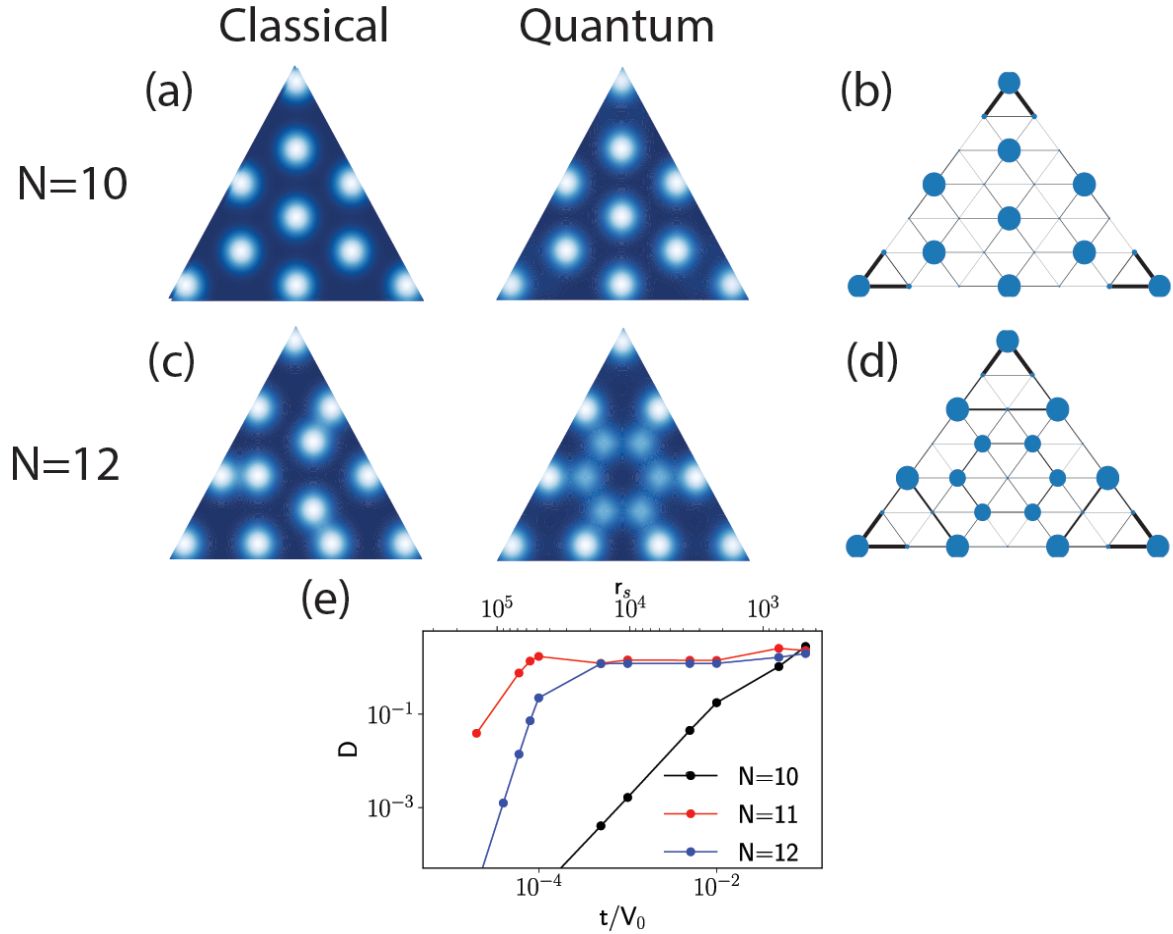

Figure 5 Density plot of the charge distribution for spin-less particles on a triangle with $L = 7$. The first row shows the *commensurate* case with $N = 10$ electrons. The second row shows the *incommensurate* case with $N = 12$ electrons. The first column represents the classical case with $t/V = 0$. The second column (a,c) shows the quantum case with $t/V = 0.01$. The size of the dot is a measure of the electron density. The last column (b,d) represents the quantum $t/V = 0.01$ solution in a graph representation where the thickness of the bond represents a relative contribution to the total kinetic energy and a size of the dot the electronic density. (e) The norm of the density difference distribution D (see the main text) for various number of electrons N versus the ratio between the hopping integral and the interaction $t/V$ or the Wigner-Seitz radius $r_s = 1.92e^2/\hbar v_F$, and $v_F$ is the Fermi velocity.

## Conclusions

In larger ETs, electron trajectories are modified by correlations within the centre, and boundary conditions at the barriers which cannot be understood solely on the basis of semi-classical correlated polaron packing, or free-electron QB. However, correlated electron modelling quite successfully predicts the appearance of domain walls in the CCDW structure forced by the confinement, which cannot be explained by any free-electron QB model. The observation of mirrored, but slightly different QPI patterns within ETs of the same size reveals that the QPI patterns are emergent, self-organized many-body electronic states. The fact that very small changes in geometry and/or barely perceptible imperfections on the atomic level give rise to dramatic changes of the QPI patterns is consistent with chaotic trajectories which originate not only from imperfect ET shapes, but also interaction with crystal lattice fluctuations and inter-particle correlations. However, the imperfections in the ET construction (Fig. 1) are expected to change the detailed QI patterns, but not the observed generic features. The understanding of behaviour and interaction of itinerant electrons with correlation-localised polarons in intertwined textures opens the way to microscopic electronics devices with correlated quantum materials. Moreover, the intertwined orders visible in the ET QPI patterns may help in understanding quantum materials in which a coexistence of itinerant and polaronic correlation-localised carriers is observed, reconciling the seeming dichotomy of different experiments that observe itinerant states (e.g. ARPES, quantum oscillations[33]) and localised states (e.g. optics[34,35], STM[36]) within the same material under seemingly identical conditions.

# Acknowledgments

We wish to acknowledge discussions with Tomaž Prosen, single crystals grown for this work by Petra Sutar, funding from the Slovenian Research Agency (ARRS), projects P1-0040, N1-0092 and young researcher grants, P17589 and P08333. D. G. acknowledges the support by ARRS under Programs No. J1-2455 and P1-0044. The Flatiron Institute is a division of the Simons Foundation. This project has received funding from the European Union's Horizon 2020 research and innovation program under the Marie Skłodowska-Curie grant agreement No 701647.

# References

1. Heller, E. J. Bound-state eigenfunctions of classically chaotic hamiltonian systems: Scars of periodic orbits. *Physical Review Letters* **53**, 1515–1518 (1984).

2. Casati, G. & Prosen, T. Mixing Property of Triangular Billiards. *Physical Review Letters* **83**, 4729–4732 (1999).

3. Marcus, C. M., Rimberg, A. J., Westervelt, R. M., Hopkins, P. F. & Gossard, A. C. Conductance fluctuations and chaotic scattering in ballistic microstructures. *Physical Review Letters* **69**, 506–509 (1992).

4. Wilkinson, P. B. *et al.* Observation of 'scarred' wavefunctions in a quantum well with chaotic electron dynamics. *Nature Nanotechnology* **380**, 608–610 (1996).

5. Linke, H., Christensson, L., Omling, P. & Lindelof, P. Stability of classical electron orbits in triangular electron billiards. *Physical Review B - Condensed Matter and Materials Physics* **56**, 1440–1446 (1997).

6. Fromhold, T. M. *et al.* Magnetotunneling spectroscopy of a quantum well in the regime of classical chaos. *Physical Review Letters* **72**, 2608–2611 (1994).

7. Ponomarenko, L., Schedin, F. & al, et. Chaotic Dirac billiard in graphene quantum dots. *Science* (2008).

8. Crook, R. *et al.* Imaging Fractal Conductance Fluctuations and Scarred Wave Functions in a Quantum Billiard. *PRL* **91**, 730–4 (2003).

9. Cabosart, A. F. N. R. A. I. S. T. S. F. and B. H. D. Recurrent Quantum Scars in a Mesoscopic Graphene Ring. *Nano Letters* **17**, 1–6 (2017).

10. Ho, W. W., Choi, S., Pichler, H. & Lukin, M. D. Periodic Orbits, Entanglement, and Quantum Many-Body Scars in Constrained Models: Matrix Product State Approach. *Phys Rev Lett* **122**, 040603 (2019).

11. Turner, C. J., Michailidis, A. A., Abanin, D. A., Serbyn, M. & Papić, Z. Weak ergodicity breaking from quantum many-body scars. *Nat Phys* **14**, 745–749 (2018).

12. Vodeb, J. *et al.* Configurational electronic states in layered transition metal dichalcogenides. *New J Phys* **21**, 083001 (2019).

13. Wilson, J. A., Disalvo, F. J. & Mahajan, S. Charge-Density Waves and Superlattices in Metallic Layered Transition-Metal Dichalcogenides. *Advances In Physics* **24**, 117–201 (1975).

14. Sipos, B., Berger, H., Forro, L., Tutis, E. & Kusmartseva, A. F. From Mott state to superconductivity in 1T-TaS2. *Nature Materials* **7**, 960–965 (2008).

15. Fazekas, P. & Tosatti, E. Charge Carrier Localization in Pure and Doped 1t-Tas2. *Physica B & C* **99**, 183–187 (1980).

16. Klanjsek, M. *et al.* A high-temperature quantum spin liquid with polaron spins. *Nature Physics* **13**, 1130–1134 (2017).

17. Karpov, P. & Brazovskii, S. Modeling of networks and globules of charged domain walls observed in pump and pulse induced states. *Scientific Reports* **8**, 1–7 (2018).

18. Hall, J., Ehlen, N., Berges, J., Loon, E. van & ACS, C. van E. Environmental Control of Charge Density Wave Order in Monolayer 2H-TaS2. *ACS Applied Nano* **13**, 10210–10220 (2019).

19. Bardeen, J. Classical versus quantum models of charge-density-wave depinning in quasi-one-dimensional metals. *Physical review. B, Condensed matter* **39**, 3528–3532 (1989).

20. Miller, J. H., Wijesinghe, A., Tang, Z. & Guloy, A. Coherent quantum transport of charge density waves. *PRB* **87**, 115127 (2013).

21. Miller, J. H., Wijesinghe, A. I., Tang, Z. & Guloy, A. M. Correlated Quantum Transport of Density Wave Electrons. *PRL* **108**, 036404 (2012).

22. Ravnik, J., Vaskivskyi, I. & Gerasimenko, Y. Strain-Induced Metastable Topological Networks in Laser-Fabricated TaS2 Polytype Heterostructures for Nanoscale Devices. *ACS Applied Nano* **2**, 3743–3751 (2019).

23. Shimada, T., Ohuchi, F. S. & Parkinson, B. A. Work function and photothreshold of layered metal dichalcogenides. *Japanese Journal of Applied Physics* **33**, 2696–2698 (1994).

24. Ma, L. *et al.* A metallic mosaic phase and the origin of Mott-insulating state in 1T-TaS2. *Nature Communications* **7**, 1–8 (2016).

25. Cho, D. *et al.* Correlated electronic states at domain walls of a Mott-charge-density-wave insulator 1 T -TaS 2. *Nature Communications* **8**, 392 (2017).

26. Panda, S., Maulik, S., Chakraborty, S. & Khastgir, S. P. From classical periodic orbits in integrable π-rational billiards to quantum energy spectrum. *The European Physical Journal Plus* **134**, 438–17 (2019).

27. Gerasimenko, Y. A., Karpov, P., Vaskivskyi, I., Brazovskii, S. & Mihailovic, D. Intertwined chiral charge orders and topological stabilization of the light-induced state of a prototypical transition metal dichalcogenide. *npj Quantum Materials* **4**, 1–9 (2019).

28. Stahl, Q. *et al.* Collapse of layer dimerization in the photo-induced hidden state of 1T-TaS2. *Nature Communications* **11**, 1–7 (2020).

29. Cho, D. *et al.* Nanoscale manipulation of the Mott insulating state coupled to charge order in 1T-TaS2. *arXiv.org* (2015).

30. Rossnagel, K. On the origin of charge-density waves in select layered transition-metal dichalcogenides. *Journal Of Physics-Condensed Matter* **23**, 213001 (2011).

31. Ryu, H. *et al.* Persistent Charge-Density-Wave Order in Single-Layer TaSe2. *ACS Applied Nano* **18**, 689–694 (2018).

32. Samajdar, R. & Jain, S. R. Nodal domains of the equilateral triangle billiard. *Journal of Physics A: Mathematical and Theoretical* **47**, 195101–23 (2014).

33. Vignolle, B., Vignolles, D., LeBoeuf, D. & Lepault, S. Quantum oscillations and the Fermi surface of high-temperature cuprate superconductors. (2011).

34. Mihailovic, D. *et al.* Application of the polaron-transport theory to sigma ( omega ) in Tl2Ba2Ca1-xGdxCu2O8, YBa2Cu3O7- d, and La2-xSrxCuO4. *Physical review. B, Condensed matter* **42**, 7989–7993 (1990).

35. Mertelj, T., Demsar, J., Podobnik, B., Poberaj, I. & Mihailovic, D. Photoexcited carrier relaxation in YBaCuO by picosecond resonant Raman spectroscopy. *PHYSICAL REVIEW B …* **55**, 6061–6069 (1997).

36. Hoffman, J. E. *et al.* Imaging quasiparticle interference in Bi2Sr2CaCu2O8+d. *Science* **297**, 1148–1151 (2002).